\documentclass[journal]{IEEEtran}
\usepackage{balance}
\vspace{-20pt}
\ifCLASSINFOpdf
\else
   \usepackage[dvips]{graphicx}
\fi
\usepackage{subfigure}
\usepackage{cite}
\usepackage{mathrsfs}
\usepackage{amsmath,amssymb,amsfonts}
\usepackage{url}
\usepackage{algorithmic}
\usepackage{lipsum}
\usepackage{multicol}
\usepackage{graphicx}
\usepackage{epstopdf}
\usepackage{textcomp}
\usepackage{caption}
\usepackage{graphicx}
\usepackage{algorithmic}
\usepackage{algorithm}
\usepackage{bm}
\usepackage{array}
\usepackage{setspace}
\usepackage{makecell,multirow,diagbox}
\usepackage{textcomp,booktabs}
\usepackage[usenames,dvipsnames]{color}
\usepackage{colortbl}
\definecolor{mygray}{gray}{.9}
\definecolor{mypink}{rgb}{.99,.91,.95}
\definecolor{mycyan}{cmyk}{.3,0,0,0}

\ifCLASSOPTIONcompsoc
\usepackage[caption=true, font=footnotesize, labelfont=sf, textfont=sf]{subfig}
\else
\usepackage[caption=true, font=footnotesize]{subfig}
\newenvironment{Proof}{{\it Proof:}}{\hfill $\blacksquare$\par}

\begin{document}
%
\title{Resource Allocation and Passive Beamforming for IRS-assisted URLLC Systems}

\author{Yangyi Zhang, Xinrong Guan, Qingqing Wu, Zhi Ji, and Yueming Cai
\vspace{-8pt}
\thanks{
Yangyi Zhang, Xinrong Guan and Zhi Ji are with the College of Communications Engineering, Army Engineering University of PLA, Nanjing, 210007, China (e-mail: zhangyy146@163.com; guanxr@ieee.org). QingqingWu is with the Department of Electronic Engineering, Shanghai JiaoTong University, Shanghai 200240, China (e-mail: qingqingwu@sjtu.edu.cn).Yueming Cai is with the Ministerial Key Laboratory of JGMT, Nanjing University of Science and Technology, Nanjing 210094 (caiym@vip.sina.com). ({\textit{Corresponding author: Xinrong Guan}}).
 }}
\maketitle
\pagestyle{empty}
\thispagestyle{empty} 

\begin{abstract}
In this correspondence, we investigate an intelligent reflective surface (IRS) assisted downlink ultra-reliable and low-latency communication (URLLC) system, where an access point (AP) sends short packets to multiple devices with the help of an IRS. Specifically, a performance comparison between the frequency division multiple access (FDMA) and time division multiple access (TDMA) is conducted for the considered system, from the perspective of average age of information (AoI). Aiming to minimize the maximum average AoI among all devices by jointly optimizing the resource allocation and passive beamforming. However, the formulated problem is difficult to solve due to the non-convex objective function and coupled variables. Thus, we propose an alternating optimization based algorithm by dividing the original problem into two sub-problems which can be efficiently solved. Simulation results show that TDMA can achieve lower AoI by exploiting the time-selective passive beamforming of IRS for maximizing the signal to noise ratio (SNR) of each device consecutively. Moreover, it also shows that as the length of information bits becomes sufficiently large as compared to the available bandwidth, the proposed FDMA transmission scheme becomes more favorable instead, due to the more effective utilization of bandwidth.  
\end{abstract}
\vspace{-4pt}
\IEEEpeerreviewmaketitle

\section{Introduction}

The strengthened ultra-reliable and low-latency communication (URLLC) proposed by 6G has more stringent requirements for reliability and latency to support services like Internet of Things (IoT) industry automation, self-driving car and telemedicine. Many of them require timely command information, for example, in IoT industry automation, command center will give command information according to its perception of current environment, and devices need to obtain latest command information for completing their works. However, due to the transmission delay, packet error, and other factors, the command information arrives with a lag time \cite{1}. To evaluate the lag time of command information, a new metric named age of information (AoI) is proposed. Specifically, AoI is the interval from the generation time of command information to the current moment, which reflects the freshness of the current command information and thus the effectiveness of the operation \cite{3}.

To reduce the transmission time, command information are usually encoded into short packets at the transmitter, which however renders increased packet error rate (PER) at the receiver due to the finite blocklength. To tackle the above challenge, \cite{5} proposed to encode messages of different users into an enlarged length packet, \cite{6} introduced the automatic repeat request (ARQ) into short packet communication system, while \cite{7} proposed to exploit collaborative relay technology to increase the signal to noise ratio (SNR) via best relay selection. However, the combination of user messages \cite{5} and retransmission of messages \cite{6} both increase the transmission delay, while the scheme based on collaborative relay \cite{7} renders additional power and hardware cost. 

Recently, intelligent reflective surface (IRS) has attracted wide attention due to its potential of smartly reconfiguring the wireless environment and thus achieving high spectrum-efficiency and energy-efficiency \cite{4}. Specifically, IRS is a uniform planar metasurface composed of a large number of passive reflecting elements. By dynamically adjusting the reflection amplitude and/or phase of each element, IRS can achieve signal enhancement and interference suppression cost-effectively \cite{8,9,10}. Thanks to such capability, IRS can also be applied to improve reliability and data rate of short packet communication systems without causing additional latency and at low cost \cite{11,12}. However, the above works only focus on IRS-aided single user case, and cannot be straightforwardly extended to the IRS-aided multiuser URLLC systems, wherein the multiple access problem becomes extremely important but still remains unsolved.

Motivated by the above, we conduct a performance comparison between the IRS assisted frequency division multiple access (FDMA) and IRS assisted time division multiple access (TDMA) for the multiuser URLLC system. Specifically, aiming to minimize the maximum average AoI, the resource allocation and passive beamforming are jointly optimized for both schemes. Note that this work is different from that in \cite{13}, which was based on the conventional infinite blocklength assumption, only considered a two-user case and simply adopted an equal time/bandwidth allocation strategy. Simulation results show that the proposed TDMA design can achieve better AoI performance than the FDMA design via dynamically tuning the passive beamforming for maximizing the SNR of each device consecutively. However, it also shows that as the available bandwidth decreases or the length of information bits increases, FDMA may outperform TDMA because in this case the bandwidth becomes the bottleneck and thus optimizing bandwidth allocation is more effective. 

\section{System model and problem formulation}

\subsection{Signal Model}

As shown in Fig. 1, we consider an IRS-assisted URLLC system where an AP intends to transmit $K$ short packets to a set of $K$ devices, respectively, with the help of an IRS consisting of $M$ reflecting elements. The baseband equivalent channels of the AP-device $k$ link, AP-IRS link and IRS-device $k$ link are denoted by ${h_{d,k}}$, ${\mathbf{h}}_{ar}^H \in {\mathbf{C}^{1 \times M}}$ and $\mathbf{h}_{rk}^H \in {\mathbf{C}^{1 \times M}}$, respectively. Assume that all channels undergo quasi-static fading, i.e., the channel coefficient remains constant within each coherence block and varies independently among different coherence blocks. To characterize the performance limit, we assume that perfect channel state information (CSI) of all links is perfectly known at the AP, e.g., by applying some efficient channel estimation methods \cite{10}. Assume that $\mathbf{\Phi } = diag\left( {{v_1},{v_2}, \cdots  \cdots ,{v_M}} \right)$ represents the diagonal phase shift matrix of the IRS, where ${v_m} = {e^{j{\theta _m}}}$, ${\theta _m} \in \left[ {0,2\pi } \right)$ represents the phase shift of the $m$-th IRS reflecting element \cite{4}. As such, the composite AP-IRS-device $k$ channel is then modeled as a concatenation of three components, namely, the AP-IRS link, IRS's reflection with phase shifts, and IRS-device $k$ link, i.e., ${{\bf{h}}_{rk}^H{\bf{\Phi h}}_{ar}^*}$. The total available bandwidth, the length of each channel coherence block, the length of information bits for each device and the power spectrum density of the noise are denoted by $B$, $T$, $D$ and $\sigma_0^2$, respectively. And the transmission power of AP is assumed to be fixed as $P$.

\begin{figure}
\begin{center}
\vspace{-0.0cm}  
\setlength{\abovecaptionskip}{0.2cm}   
\setlength{\belowcaptionskip}{-0.6cm}   
  \includegraphics[width=3.2in]{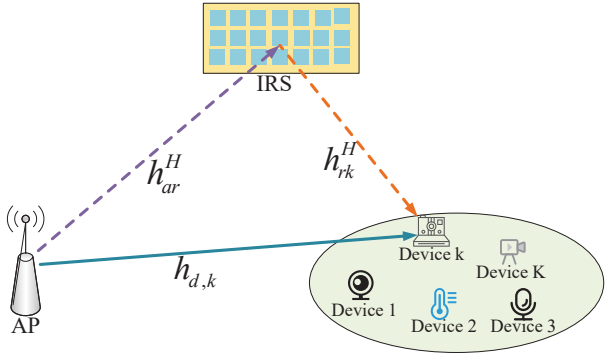}\\
  \caption{IRS-assisted URLLC system.}\label{Fig:model}
\end{center}
\end{figure}
\subsection{FDMA Transmission Scheme}

In FDMA transmission scheme, the AP communicates with the $K$ devices simultaneously in each channel coherence block. Denoting the bandwidth and the power allocated to device $k$ by ${B_k}$ and ${P_{F{\rm{k}}}}$, the signal received by device $k$ is expressed as
\begin{equation}\label{1}
{y_{Fk}} = \left( {{h_{d,k}} + {\bf{h}}_{rk}^H{\bf{\Phi h}}_{ar}^*} \right)\sqrt {{P_{Fk}}} {s_k} + {n_{Fk}},
\end{equation}
where ${s_k} \sim CN\left( {0,1} \right)$ is the signal transmitted from AP to device $k$, ${n_{Fk}} \sim CN\left( {0,{B_k}\sigma _0^2} \right)$ is the complex additive white Gaussian noise (AWGN). By denoting ${\mathbf{h}_{ark}} = diag\left( {\mathbf{h}_{rk}^H} \right)\mathbf{h}_{ar}^*$ and ${\mathbf{v}^H} = \left[ {{v_1},{v_2}, \cdots  \cdots ,{v_M}} \right]$, we have $\mathbf{h}_{rk}^H\mathbf{\Phi} \mathbf{h}_{ar}^* = {\mathbf{v}^H}{\mathbf{h}_{ark}}$. Thus, the SNR of device $k$ is expressed as
\vspace{-4pt}
\begin{equation}\label{2}
{\gamma _{Fk}} = \frac{{{P_{F{\rm{k}}}}{{\left| {{h_{d,k}} + {\mathbf{v}^H}{\mathbf{h}_{ark}}} \right|}^2}}}{{{B_k}\sigma _0^2}}.
\end{equation}

The $D$ bits information is encoded into a short packet with the blocklength of $B_kT$, thus the packet error rate at device $k$ can be expressed as \cite{14}
\vspace{-4pt}

\begin{equation}\label{3}
{\varepsilon _{Fk}} = Q\left[ {\frac{{\ln \left( {1 + {\gamma _{Fk}}} \right) - \ln 2\frac{D}{{{B_k}T}}}}{{\sqrt {1 - {{\left( {1 + {\gamma _{Fk}}} \right)}^{ - 2}}} /\sqrt {{B_k}T} }}} \right],
\end{equation}
where $Q\left( x \right) = \int_x^{ + \infty } {\frac{1}{{\sqrt {2\pi } }}} \exp \left( { - \frac{1}{2}{t^2}} \right)dt$ is complementary cumulative distribution function.

We consider the generate-at-will scheme for the generation of short packets for each device \cite{15}. Assuming that AP generates packets immediately at the beginning of each coherence block, AoI refers to the interval from generation time of the latest valid packet to present moment.
As shown in Fig. 2(a), taking the device 1 for example, AoI increases linearly over time until it successfully decodes a short packet. At this point, AoI is reset to $T$, as the latest valid packet was generated one coherence block ago.



Considering a long-term average of information freshness, average AoI is stable and suitable as a performance metric. According to \cite{16}, the average AoI of device $k$ in FDMA scheme can be expressed as
\vspace{-4pt}

\begin{equation}\label{6}
{\overline \Delta  _{Fk}} = \frac{T}{2}\left( {\frac{2}{{1 - {\varepsilon _{Fk}}}} + 1} \right).
\end{equation}

Our goal is to minimize the maximum average AoI among all devices by jointly optimizing the power/bandwidth allocation and passive beamforming. Thus, the optimization problem can be formulated as

\begin{subequations}\label{eq:P1}
\begin{spacing}{0.3}
\begin{flalign}\label{eq:P7a}
\:\:\:\:\:\:({\rm{\mathbf{P1}}}):&\mathop {\min }\limits_{{{\rm{B}}_k},{{\rm{P}}_{Fk}},{\bf{v}}} \max \left\{ {{{\bar \Delta }_{Fk}}\left| {k = 1,2, \cdots  \cdots ,K} \right.} \right\},&
\end{flalign}
\begin{flalign}\label{eq:P7b}
&\:\:\:\:\:\:\:\:\:\:\:\:\:\:\:\:\:\:\:\:\:\:\:\:\:\:\:\:{\rm{s}}{\rm{.t}}{\rm{.}}\sum\limits_{k = 1}^K {{B_k}}  = B,&
\end{flalign}
\begin{flalign}\label{eq:P7c}
&\:\:\:\:\:\:\:\:\:\:\:\:\:\:\:\:\:\:\:\:\:\:\:\:\:\:\:\:\:\:\:\:\:\: \sum\limits_{k = 1}^K {{P_{Fk}}}  = P,&
\end{flalign}
\begin{flalign}\label{eq:P7d}
&\:\:\:\:\:\:\:\:\:\:\:\:\:\:\:\:\:\:\:\:\:\:\:\:\:\:\:\:\:\:\:\:\:\: \left| {{v_m}} \right| = 1,{\kern 1pt} {\kern 1pt} {\kern 1pt} {\kern 1pt} {\kern 1pt} {\kern 1pt} {\kern 1pt} {\kern 1pt} {\kern 1pt} {\kern 1pt} {\kern 1pt} {\kern 1pt} \forall m.&
\end{flalign}
\end{spacing}
\end{subequations}

\begin{figure*}[t]
	\centering
	\subfigure[{}]{
		\begin{minipage}{0.4\linewidth}
			\centering
			\includegraphics[width=3.0in,height=2in]{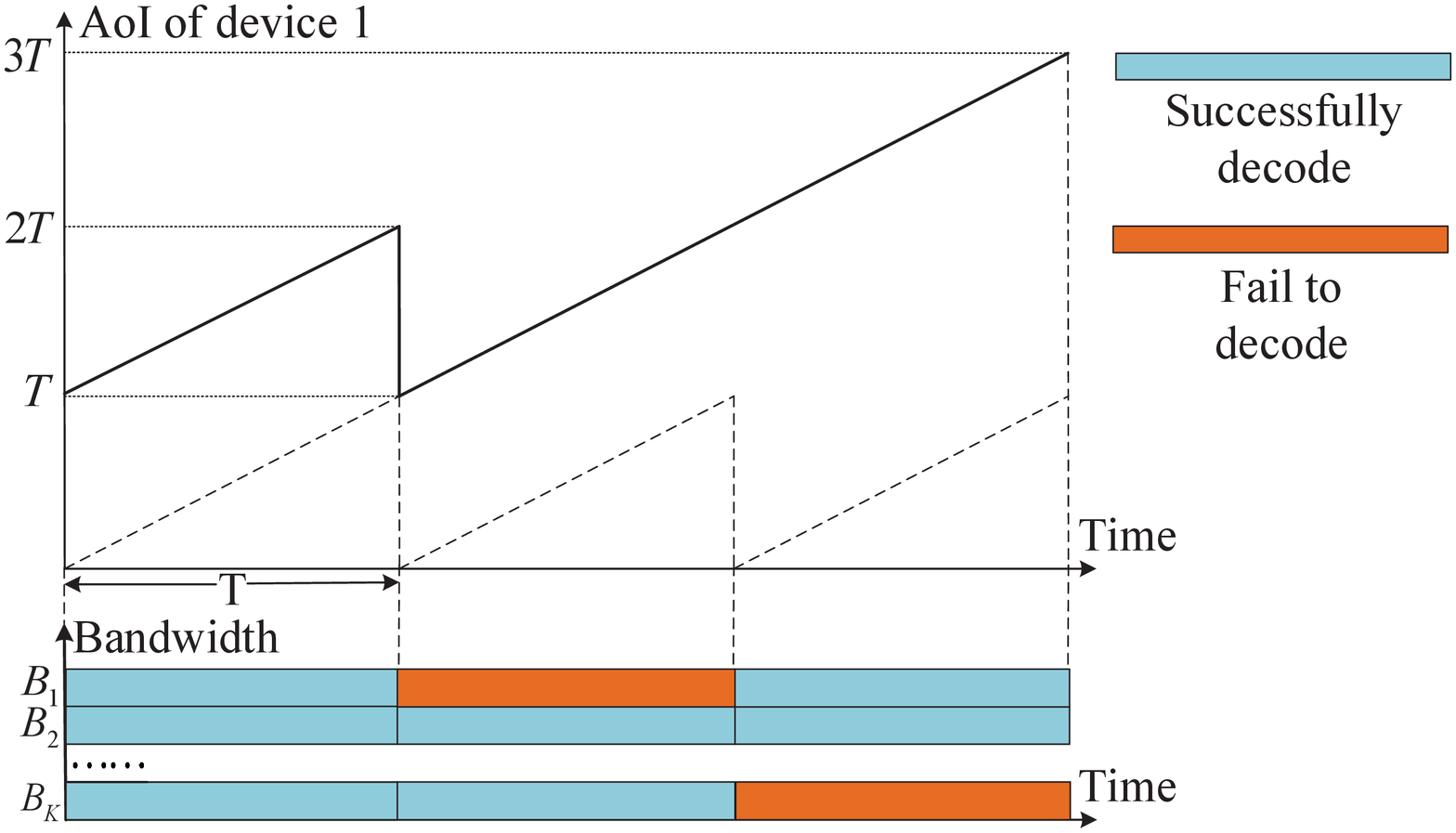}
		\end{minipage}
	} \hspace{3mm}
	\subfigure[{}]{
		\begin{minipage}{0.5\linewidth}
			\centering
			\includegraphics[width=3.4in,height=2in]{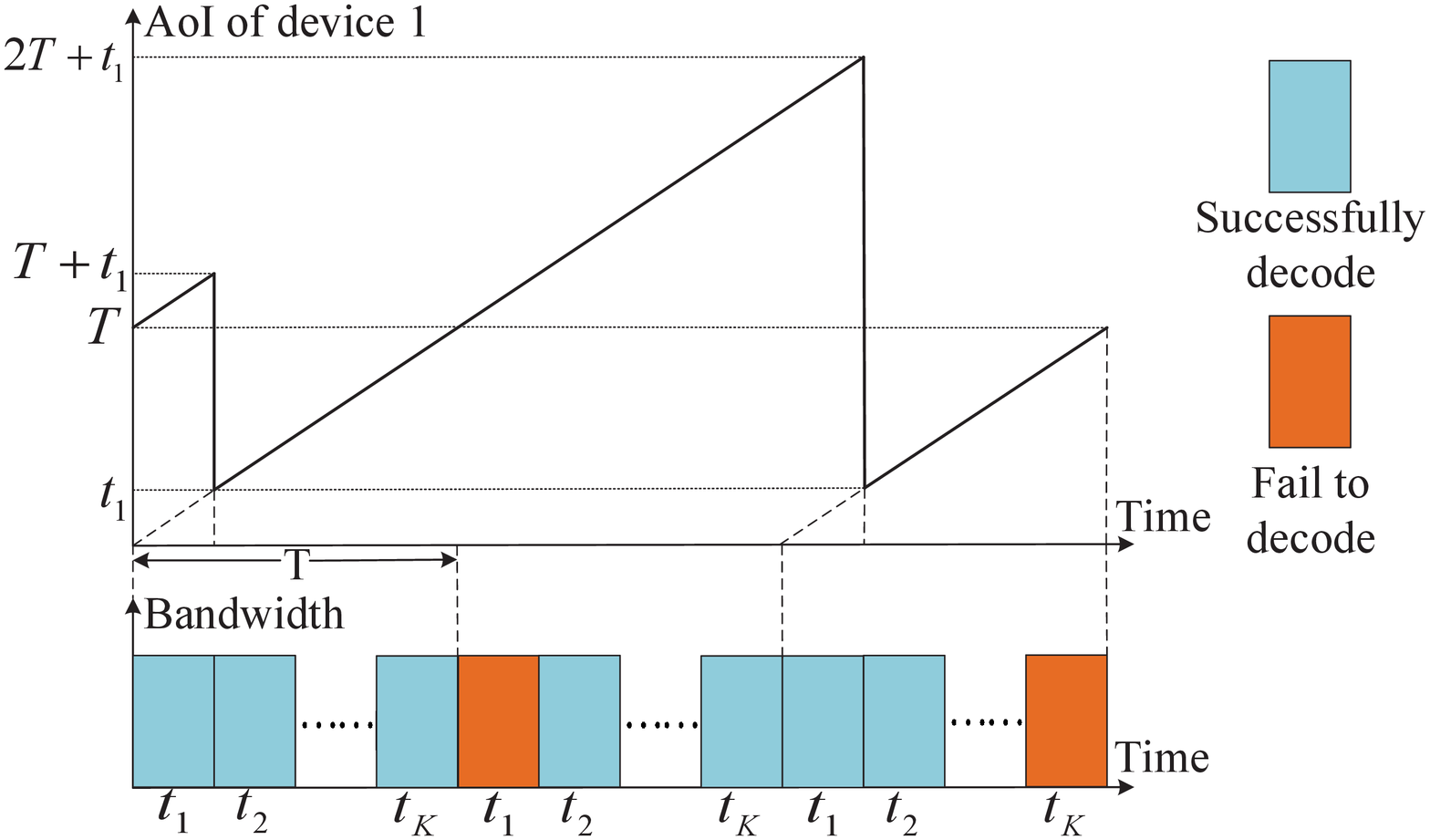}
		\end{minipage}
	}
	\caption{AoI evolution for (a) FDMA scheme and (b) TDMA scheme.}
	\label{fig2}
	\vspace{-5mm}
\end{figure*}

\vspace{4pt}

\subsection{TDMA Transmission Scheme}

In TDMA transmission scheme, the AP sends short packets to each device consecutively in each coherence block, while the time slot allocated to device $k$ is denoted by ${t_k}$. Thus, the signal received by device $k$ is given by
\vspace{-4pt}

\begin{equation}\label{8}
{y_{Tk}} = \left( {{h_{d,k}} + {\bf{h}}_{rk}^H{\bf{\Phi h}}_{ar}^*} \right)\sqrt P {s_k} + {n_{Tk}},
\end{equation}
where ${n_{Tk}} \sim CN\left( {0,B\sigma _0^2} \right)$. Correspondingly, the SNR and the packet error rate can be respectively expressed as
\vspace{-4pt}
\begin{equation}\label{9}
{\gamma _{Tk}} = \frac{{P{{\left| {{h_{d,k}} + {\mathbf{v}^H}{\mathbf{h}_{ark}}} \right|}^2}}}{{{B\sigma _0^2}}},
\end{equation}
\begin{equation}\label{10}
{\varepsilon _{Tk}} = Q\left[ {\frac{{\ln \left( {1 + {\gamma _{Tk}}} \right) - \ln 2\frac{D}{{B{t_k}}}}}{{\sqrt {1 - {{\left( {1 + {\gamma _{Tk}}} \right)}^{ - 2}}} /\sqrt {B{t_k}} }}} \right].
\end{equation}

AoI evolution in TDMA scheme is as shown in Fig. 2(b). By denoting ${T_k} = \sum\limits_{j = 1}^k {{t_j}} $, if device $k$ successfully decodes a short packet, corresponding AoI will be reset to ${T_k}$ as it takes a time interval of ${T_k}$ since the beginning of the latest coherence block. Similarly, the average AoI of device $k$ in TDMA scheme can be expressed as

\begin{equation}\label{11}
{\overline \Delta  _{Tk}} = \frac{T}{2}\left( {\frac{2}{{1 - {\varepsilon _{Tk}}}} - 1} \right) + {T_k}.
\end{equation}
As a result, the optimization problem is formulated as
\vspace{-4pt}

\begin{subequations}\label{eq:P2}
\begin{spacing}{0.3}
\begin{flalign}\label{eq:P12a}
\:\:\:\:\:\:\:\:\:({\rm{\mathbf{P2}}}):&\mathop {\min }\limits_{{{\rm{t}}_k},{\bf{v}}} \max \left\{ {{{\bar \Delta }_{Tk}}\left| {k = 1,2, \cdots  \cdots ,K} \right.} \right\},&
\end{flalign}
\begin{flalign}\label{eq:P12b}
&\:\:\:\:\:\:\:\:\:\:\:\:\:\:\:\:\:\:\:\:\:\:\:{\rm{s}}{\rm{.t}}{\rm{.}}\sum\limits_{k = 1}^K {{t_k}}  = T,&
\end{flalign}
\begin{flalign}\label{eq:P12c}
&\:\:\:\:\:\:\:\:\:\:\:\:\:\:\:\:\:\:\:\:\:\:\:\:\:\:\:\: \left| {{v_m}} \right| = 1,{\kern 1pt} {\kern 1pt} {\kern 1pt} {\kern 1pt} {\kern 1pt} {\kern 1pt} {\kern 1pt} {\kern 1pt} {\kern 1pt} {\kern 1pt} {\kern 1pt} {\kern 1pt} \forall m.&
\end{flalign}
\end{spacing}
\end{subequations}
\vspace{10pt}

Both (P1) and (P2) are difficult to solve due to the non-convex objective functions and coupled variables. However, we notice that the resultant problems can be effectively solved when one of the resource allocation and passive beamforming are fixed. Therefore, we propose an alternating optimization based algorithm to solve (P1) and (P2), respectively.


\section{Proposed Solutions}


\subsection{FDMA Transmission Design}

It can be seen from Eq. (4) that average AoI ${\overline \Delta  _{Fk}}$ increases as the packet error rate ${{\varepsilon _{Fk}}}$. Furthermore, the packet error rate ${{\varepsilon _{Fk}}}$ in Eq. (3) is a complementary cumulative distribution function and $Q\left( x \right)$ is a decreasing function of $x$. So ${\overline \Delta  _{Fk}}$ goes down as ${\frac{{\ln \left( {1 + {\gamma _{Fk}}} \right) - \ln 2\frac{D}{{{B_k}T}}}}{{\sqrt {1 - {{\left( {1 + {\gamma _{Fk}}} \right)}^{ - 2}}} /\sqrt {{B_k}T} }}}$ goes up. Since we have $\sqrt {1 - {{\left( {1 + {\gamma _{Fk}}} \right)}^{ - 2}}}  \approx 1$, (P1) can be approximated as

\begin{subequations}\label{eq:P3}
\addtolength{\jot}{-1mm}
\small
\vspace{-1mm}
\begin{spacing}{0.5}
\begin{flalign}\label{eq:P13a}
({\rm{\mathbf{P3.1}}})\!:\!&\mathop {\max }\limits_{\scriptstyle{{\rm{B}}_k}\!,\!{\bf{v}}\!,\!\hfill\atop
\scriptstyle{\rm{ }}{{\rm{P}}_{Fk}}\hfill} \min \left\{\! {\sqrt {{B_k}T} \ln \left( {1 \!+\! {\gamma _{Fk}}} \right) \!-\! \ln 2\frac{D}{{\sqrt {{B_k}T} }}\left| {\forall k} \right.}\! \right\}\!,\!&
\end{flalign}
\begin{flalign}\label{eq:P13b}
&\:\:\:\:\:\:\:\:\:\:\:\:\:\:\:\:\:\:{\rm{s}}{\rm{.t}}{\rm{.}}(5b)-(5d).&
\end{flalign}
\end{spacing}
\end{subequations}
\vspace{7pt}

\subsubsection{Optimizing $\left\{ {{B_k}} \right\}$ for given $\left\{ {{P_{Fk}}} \right\}$ and $\mathbf{v}$}
By \negthinspace denoting \negthinspace ${f_k}\left( {{B_k}} \right) \!=\! \sqrt {{B_k}T} \ln \left( {1\! +\! {\gamma _{Fk}}} \right) \!-\! \ln 2\frac{D}{{\sqrt {{B_k}T} }}$, it \negthinspace can \negthinspace be \negthinspace observed \negthinspace that \negthinspace ${f_k}\left( {{B_k}} \right)$ \negthinspace increases \negthinspace with \negthinspace ${{B_k}}$, \negthinspace thus \negthinspace (P3.1) \negthinspace can \negthinspace be \negthinspace converted \negthinspace to

\begin{subequations}\label{eq:P3}
\begin{spacing}{0.3}
\begin{flalign}\label{eq:P14a}
\:\:\:\:\:\:\:\:\:\:\:\:\:\:\:\:\:\:\:\:\:\:({\rm{\mathbf{P3.2}}}):&\mathop {\max }\limits_{{{{B_k}}}} \min \left\{ {{f_k}\left( {{B_k}} \right)\left| {\forall k} \right.} \right\},&
\end{flalign}
\begin{flalign}\label{eq:P14b}
&\:\:\:\:\:\:\:\:\:\:\:\:\:\:\:\:\:\:\:\:\:\:\:\:\:\:\:\:\:\:\:\:\:\:\:\:\:\:\:\:\:{\rm{s}}{\rm{.t}}{\rm{.}}\sum\limits_{k = 1}^K {{B_k}}  = B.&
\end{flalign}
\end{spacing}
\end{subequations}
\vspace{10pt}
\noindent
To solve (P3.2), we resort to the following lemma.

\textbf{Lemma 1.} \emph{Assume that $\left\{ {B_k^*} \right\}$ are the optimal solutions of (P3.2), then there have ${f_1}\left( {B_1^*} \right) = {f_2}\left( {B_2^*} \right) =  \cdots    = {f_K}\left( {B_K^*} \right)$}.

\begin{Proof}
The equivalence can be verified by contradiction. If the equation in Theorem 1 did not work, then we have $\min \left\{ {{f_k}\left( {B_k^*} \right)\left| {\forall k} \right.} \right\} \ne \max \left\{ {{f_k}\left( {B_k^*} \right)\left| {\forall k} \right.} \right\}$. By denoting ${f_i}\left( {B_i^*} \right) = \min \left\{ {{f_k}\left( {B_k^*} \right)\left| {\forall k} \right.} \right\}$ and ${f_j}\left( {B_j^*} \right) = \max \left\{ {{f_k}\left( {B_k^*} \right)\left| {\forall k} \right.} \right\}$, we can always find an $\Delta B$ satisfying ${f_i}\left( {B_i^* + \Delta B} \right) = {f_j}\left( {B_j^* - \Delta B} \right)$. It is easy to know that the scheme after adjusting the bandwidth is better than the original scheme, thus the hypothesis is not valid.
\end{Proof}
By denoting ${f_1}\left( {{B_1}} \right) = {f_2}\left( {{B_2}} \right) =  \cdots  = {f_K}\left( {{B_K}} \right) = \Lambda $, then we have ${B_k} = f_k^{ - 1}\left( \Lambda  \right)$. Substitute it into the constraint (12b), the optimized bandwidth allocation is then obtained.

\subsubsection{Optimizing $\left\{ {{P_{Fk}}} \right\}$ for given $\left\{ {{B_k}} \right\}$ and $\mathbf{v}$}
 Let ${g_k}\left( {{P_{Fk}}} \right) = \sqrt {{B_k}T} \ln \left( {1 + {\gamma _{Fk}}} \right) - \ln 2\frac{D}{{\sqrt {{B_k}T} }}$, we can find that ${g_k}\left( {{P_{Fk}}} \right)$ increases with ${{P_{Fk}}}$. Similar to the optimization of $\left\{ {{B_k}} \right\}$, the optimal $\left\{ {{P_{Fk}}} \right\}$ can be obtained.

\subsubsection{Optimizing $\mathbf{v}$ for given $\left\{ {{B_k}} \right\}$ and $\left\{ {{P_{Fk}}} \right\}$}
For \negthinspace ease \negthinspace of \negthinspace exposition, we \negthinspace rewrite \negthinspace ${\left| {{h_{d,k}} \!+\! {{\bf{v}}^H}{{\bf{h}}_{ark}}} \right|^2}$ \negthinspace as \negthinspace ${\left| {{h_{d,k}}} \right|^2}\! + \!tr\left( \bf{RV} \right)$, where \negthinspace $\mathbf{R}\! =\! \left[ {\begin{array}{*{20}{c}}
{{\mathbf{h}_{ark}}\mathbf{h}_{ark}^H}&{{\mathbf{h}_{ark}}h_{d,k}^H}\\{{h_{d,k}}\mathbf{h}_{ark}^H}&0\end{array}} \right]$, $\mathbf{V}  \!= \!\widetilde v{\widetilde v^H}$, $\widetilde v \!=\! \left[ {\begin{array}{*{20}{c}}\mathbf{v}\\t\end{array}} \right]$ and $\left| t \right| \!=\! 1$ is an auxiliary variable. Note that it follows \negthinspace $\mathbf{V}  \underset{\raise0.3em\hbox{$\smash{\scriptscriptstyle-}$}}{\succ } 0$ \negthinspace and \negthinspace rank$\left( {\bf{V}} \right) \!=\! 1$. Since \negthinspace the \negthinspace rank-1 \negthinspace constraint is non-convex, we \negthinspace apply \negthinspace semidefinite \negthinspace relaxation \negthinspace (SDR) to relax this \negthinspace constraint. Specifically, by \negthinspace defining \negthinspace ${h_k}\left( {\bf{V}} \right) \!=\! \sqrt {{B_k}T} \ln \left( {1 \!+\! \frac{{{P_{F{\rm{k}}}}\left( {{{\left| {{h_{d,k}}} \right|}^2} \!+\! tr\left( \bf{RV} \right)} \right)}}{{{B_k}\sigma _0^2}}} \right) \!-\! \ln 2\frac{D}{{\sqrt {{B_k}T} }}$, (P3.1) \negthinspace can \negthinspace be converted \negthinspace to
\begin{subequations}\label{eq:P3}
\begin{spacing}{0.15}
\begin{flalign}\label{eq:P15a}
\:\:\:\:\:\:\:\:\:\:({\rm{\mathbf{P3.3}}}):& \mathop {\max }\limits_{\bf{V}} \min \left\{ {{h_k}\left( {\bf{V}} \right)\left| {\forall k} \right.} \right\},&
\end{flalign}
\begin{flalign}\label{eq:P15b}
&\:\:\:\:\:\:\:\:\:\:\:\:\:\:\:\:\:\:\:\:\:\:\:\:\:\:\:\:\:\:{\rm{s}}{\rm{.t}}{\rm{.}} \mathbf{V}  \underset{\raise0.3em\hbox{$\smash{\scriptscriptstyle-}$}}{\succ } 0,&
\end{flalign}
\begin{flalign}\label{eq:P15b}
&\:\:\:\:\:\:\:\:\:\:\:\:\:\:\:\:\:\:\:\:\:\:\:\:\:\:\:\:\:\:\:\:\:\:\:\: {{\bf{V}}_{m,m}} = 1,m = 1, \cdots ,M + 1.&
\end{flalign}
\end{spacing}
\end{subequations}
\vspace{14pt}
\noindent
Introducing a slack variable $\chi $, (P3.3) can be equivalently written as
\vspace{-8pt}
\begin{subequations}\label{eq:P3}
\begin{spacing}{0.5}
\begin{flalign}\label{eq:P18a}
\:\:\:\:\:\:\:\:\:\:\:\:\:\:\:\:\:\:\:\:\:\:\:\:({\rm{\mathbf{P3.4}}}):&\mathop {\max }\limits_{{\bf{\bf{V} }}} \chi ,&
\end{flalign}
\begin{flalign}\label{eq:P18b}
&\:\:\:\:\:\:\:\:\:\:\:\:\:\:\:\:\:\:\:\:\:\:\:\:\:\:\:\:\:\:\:\:\:\:\:\:\:\:\:\:\:\:\:{\rm{s}}{\rm{.t}}{\rm{.}} {h_k}\left( {\bf{V}} \right) \ge \chi   , {\kern 1pt} {\kern 1pt} {\kern 1pt} {\kern 1pt} {\kern 1pt}\forall k, &
\end{flalign}
\begin{flalign}\label{eq:P13b}
&\:\:\:\:\:\:\:\:\:\:\:\:\:\:\:\:\:\:\:\:\:\:\:\:\:\:\:\:\:\:\:\:\:\:\:\:\:\:\:\:\:\:\:\:\:\:\:\:\:(13b),(13c).&
\end{flalign}
\end{spacing}
\end{subequations}
\vspace{8pt}
\noindent
Given a specific value for $\chi $, (P3.4) is a standard convex semidefinite program (SDP) and can be solved by CVX. In order to find the maximum $\chi $ that satisfies the constraint (14b), the bisection search algorithm is adopted, which starts with the minimum value 0 and maximum value $\bar \chi  = \max \left\{ {\sqrt {{B_k}T} \ln \left( {1 + \frac{{{P_{F{\rm{k}}}}{{\left( {\left| {{h_{d,k}}} \right| + \left| {{{\bf{h}}_{ark}}} \right|} \right)}^2}}}{{{B_k}\sigma _0^2}}} \right) - \ln 2\frac{D}{{\sqrt {{B_k}T} }}\left| {\forall k} \right.} \right\}$. A set of lower half or upper half interval is selected recursively based on CVX checking for further search until the interval length is tolerated.

\subsubsection{Overall Algorithm}
The overall algorithm for solving (P1) is given in Algorithm 1. As the objective value is not reduced in each iteration optimization process and there is an exact upper bound, Algorithm 1 always converges. If the obtained $\mathbf{V}$ is not of rank-1, we can get an approximation of $\mathbf{v}$ by Gaussian randomization as in \cite{4}.
\begin{algorithm}[t]

	\renewcommand{\algorithmicrequire}{\textbf{Input:}}
	\renewcommand{\algorithmicensure}{\textbf{Output:}}
	\caption{An alternating algorithm for solving (P1)}
	\begin{algorithmic} [1]
		\STATE\textbf{Initialization:}
		Initialize \negthinspace $i = 0$ \negthinspace and \negthinspace input \negthinspace $\left\{ {{{\left\{ {{B_k}} \right\}}^{\left( i \right)}},{{\left\{ {{{\rm{P}}_{Fk}}} \right\}}^{\left( i \right)}},{{\bf{V}}^{\left( i \right)}}} \right\}$.
		\STATE\textbf{Repeat}
		\STATE\quad For given ${{{\left\{ {{{\rm{P}}_{Fk}}} \right\}}^{\left( i \right)}}}$ and ${{{\bf{V}}^{\left( i \right)}}}$, optimize ${\left\{ {{B_k}} \right\}^{\left( {i + 1} \right)}}$.
        \STATE\quad For given ${\left\{ {{B_k}} \right\}^{\left( {i + 1} \right)}}$ and ${{{\bf{V}}^{\left( i \right)}}}$, optimize ${{{\left\{ {{{\rm{P}}_{Fk}}} \right\}}^{\left( i +1 \right)}}}$.
        \STATE\quad For given ${\left\{ {{B_k}} \right\}^{\left( {i + 1} \right)}}$ and ${{{\left\{ {{{\rm{P}}_{Fk}}} \right\}}^{\left( i +1 \right)}}}$, optimize ${{{\bf{V}}^{\left( i+1 \right)}}}$.
		\STATE\quad Let $i \leftarrow i+1$.
		\STATE{\textbf{Until} \negthinspace the \negthinspace fractional \negthinspace increase \negthinspace of \negthinspace $\max \left\{ {{{\overline \Delta  }_{Fk}}\left| {\forall k} \right.} \right\}$ \negthinspace is \negthinspace below \negthinspace a \negthinspace threshold.}	
	\end{algorithmic}
\end{algorithm}

\subsection{TDMA Transmission Design}

In TDMA scheme, each channel coherence block is divided into $K$ time slots for devices, in each of which the AP transmits a short packet to the serving device with the help of IRS. Note that in this transmission mode, the reflecting coefficients of IRS elements are dynamically tuned during each coherence block for maximizing the SNR at each device consecutively (the so-called time-selective passive beamforming as in \cite{13}), unlike remaining unchanged in the FDMA case.

\subsubsection{Optimizing $\mathbf{v}$ for (P2)}
It can be seen from (8) and (9) that ${\overline \Delta  _{Tk}}$ decreases with the increase of ${\gamma _{Tk}}$. To maximize the ${\gamma _{Tk}}$, the passive beamforming vector of device $k$ is given by
\begin{equation}\label{19}
\mathbf{v}_k^H = \left[ {{v_{k1}},{v_{k2}}, \cdots  \cdots ,{v_{kM}}} \right],
\end{equation}
where ${v_{km}} = {e^{j{\theta _{km}}}}$, ${\theta _{km}} = \angle {h_{d,k}} - \angle h_{ark}^{\left( m \right)}$, $\angle {h_{d,k}}$ is the phase of ${h_{d,k}}$ and $\angle h_{ark}^{\left( m \right)}$ is the phase of the $m$-th element of ${{\mathbf{h}_{ark}}}$. As such,
\vspace{-8pt}
\begin{equation}\label{20}
\gamma _{Tk}^{\max } = \frac{{P{{\left( {\left| {{h_{d,k}}} \right| + \sum\limits_{m = 1}^M {\left| {h_{ark}^{\left( m \right)}} \right|} } \right)}^2}}}{{B\sigma _0^2}}.
\end{equation}

\subsubsection{Optimizing $\left\{ {{t_k}} \right\}$ for (P2)}
As (P2) is non-convex, it is intractable to solve the problem directly. Thus, we use the principle of selecting the best to fill the weakness. Specifically, by setting ${t_1} = {t_2} =  \cdots  \cdots  = {t_K} = T/K$, the initialized $\left\{ {{{\overline \Delta  }_{Tk}}\left| {\forall k} \right.} \right\}$ are thus obtained. We assume that the $i$-th and the $j$-th device achieves the maximum and minimum average AoI, respectively, i.e., ${\overline \Delta  _{Ti}} = \max \left\{ {{{\overline \Delta  }_{Tk}}\left| {\forall k} \right.} \right\}$ and ${\overline \Delta  _{Tj}} = \min \left\{ {{{\overline \Delta  }_{Tk}}\left| {\forall k} \right.} \right\}$. Then, find a $\Delta T$ such that ${t_i} = {t_i} + \Delta T$, ${t_j} = {t_j} - \Delta T$, and thus ${\overline \Delta  _{Ti}} = {\overline \Delta  _{Tj}}$. Repeat the above procedure until the maximum average AoI no longer reduced. The detailed algorithm is summarized as in Algorithm 2.
\begin{algorithm}[t]

	\renewcommand{\algorithmicrequire}{\textbf{Input:}}
	\renewcommand{\algorithmicensure}{\textbf{Output:}}
	\caption{Optimizing $\left\{ {{t_k}} \right\}$ algorithm for solving (P2)}
	\begin{algorithmic} [1]
		\STATE\textbf{Initialization:}
		Initialize $l = 0$ and input ${\left\{ {{t_k}} \right\}^{\left( l \right)}}$.
		\STATE\textbf{Repeat}
		\STATE\quad Calculate ${\left\{ {{{\overline \Delta  }_{Tk}}\left| {\forall k} \right.} \right\}^{\left( l \right)}}$, find the device ${i^{\left( l \right)}}$ whose
average AoI is $\max {\left\{ {{{\overline \Delta  }_{Tk}}\left| {\forall k} \right.} \right\}^{\left( l \right)}}$ and the device ${j^{\left( l \right)}}$ whose
average AoI is $\min {\left\{ {{{\overline \Delta  }_{Tk}}\left| {\forall k} \right.} \right\}^{\left( l \right)}}$.
		\STATE\quad Find a $\Delta {T^{\left( l+1 \right)}}$ such that $t_i^{\left( l+1 \right)} = t_i^{\left( l+1 \right)} + \Delta {T^{\left( l+1 \right)}}$, $t_j^{\left( l+1 \right)} = t_j^{\left( l+1 \right)} - \Delta {T^{\left( l+1 \right)}}$, and thus $\overline \Delta  _{Ti}^{\left( l+1 \right)} = \overline \Delta  _{Tj}^{\left( l+1 \right)}$.
		\STATE\quad Let $l \leftarrow l+1$.
		\STATE{\textbf{Until} the fractional decrease of $\max \left\{ {{{\bar \Delta }_{Tk}}\left| {\forall k} \right.} \right\}$ is below a threshold.}
	\end{algorithmic}
\end{algorithm}

\section{Simulation results}

 Unless otherwise specified, the parameters are set as follows: the AP and the IRS are located at (0, 0) and (120, 30) in meter (m), respectively, while the $K = 5$ devices are randomly distributed in a circle with the centre and radius are set as (120, 0) and $R = 10$ m. The channel between arbitrary two nodes is represented by $h = \sqrt {{L_0}{d^{ - \alpha }}} g$, where ${L_0} =  - 30$ dB is path loss at the reference distance $d_0=1$ m, $d$ denotes the distance, $\alpha$ denotes the corresponding path loss exponent and $g$ is the small-scale fading component. Considering that IRS can be flexibly deployed at a place with less scattering environment and thus we set ${\alpha _1} = 3.5$ and ${\alpha _2} = {\alpha _3} = 2.5$ for the path loss exponent of AP-device link, AP-IRS link and IRS-device link, respectively. We assume that the transmission power of the AP is $P = 0 $ dBm, the number of reflecting elements in IRS is $M = 80$, while the total available bandwidth, the length of each channel coherence block, the length of information bits for each device and the power spectrum density of the noise are set as $B = 1$ MHz, $T = 1$ ms, $D = 600$ bits and $\sigma _0^2 =  - 168$ dBm/Hz, respectively. In addition to the proposed two designs (FDMA, Proposed) and (TDMA, Proposed), the equal time/bandwidth allocation strategy in \cite{13} (FDMA, Equal \cite{13}) and (TDMA, Equal \cite{13}) are also adopted for performance comparison.
 \vspace{-10pt}

\begin{figure*}[t]
	\centering
	\subfigure[{The maximum average AoI versus the number of reflecting elements.}]{
		\begin{minipage}{0.3\linewidth}
			\centering
			\includegraphics[width=2.22in,height=1.93in]{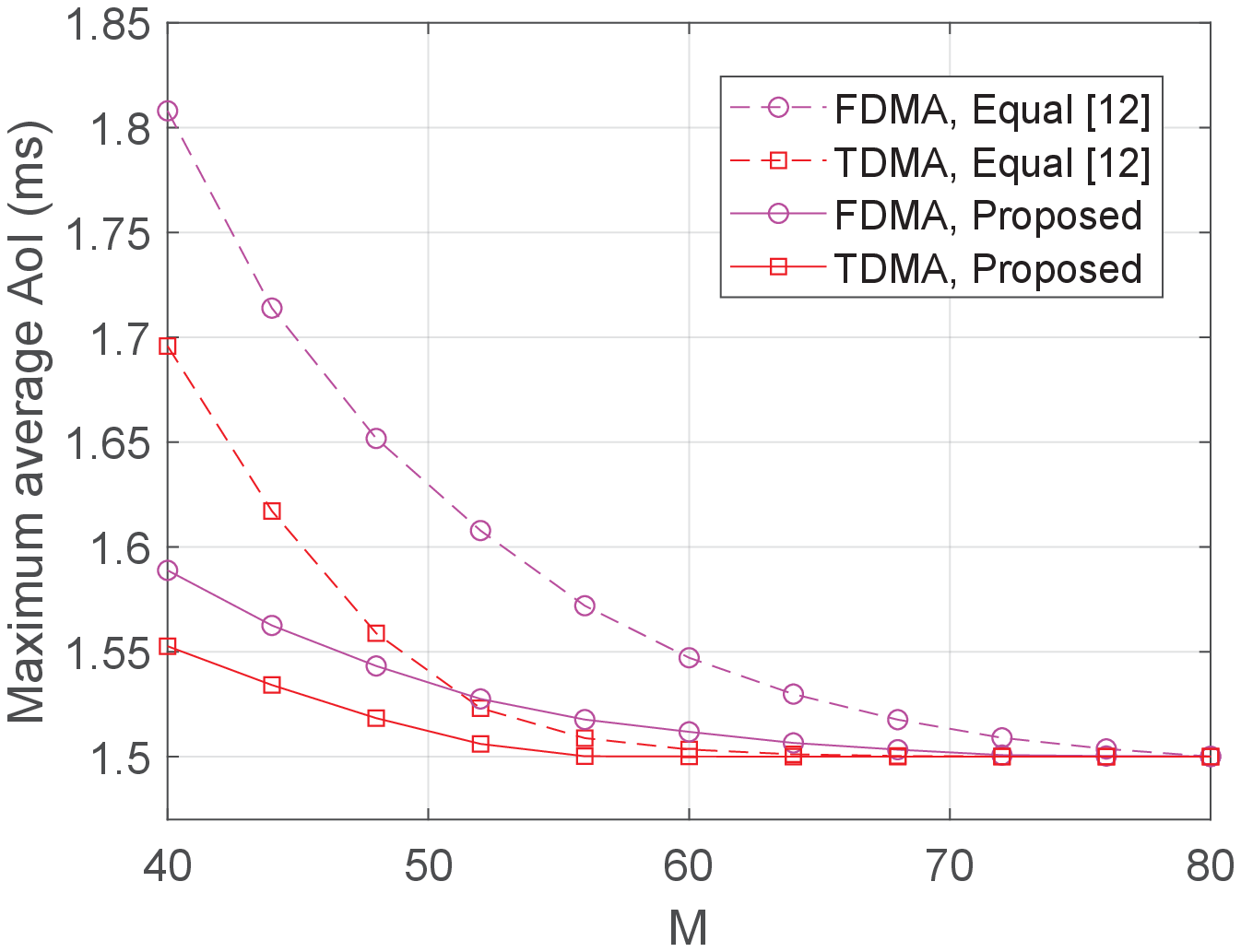}
		\end{minipage}
	}  \hspace{3mm}
	\subfigure[{The maximum average AoI versus the total available bandwidth.}]{
		\begin{minipage}{0.3\linewidth}
			\centering
			\includegraphics[width=2.22in,height=1.93in]{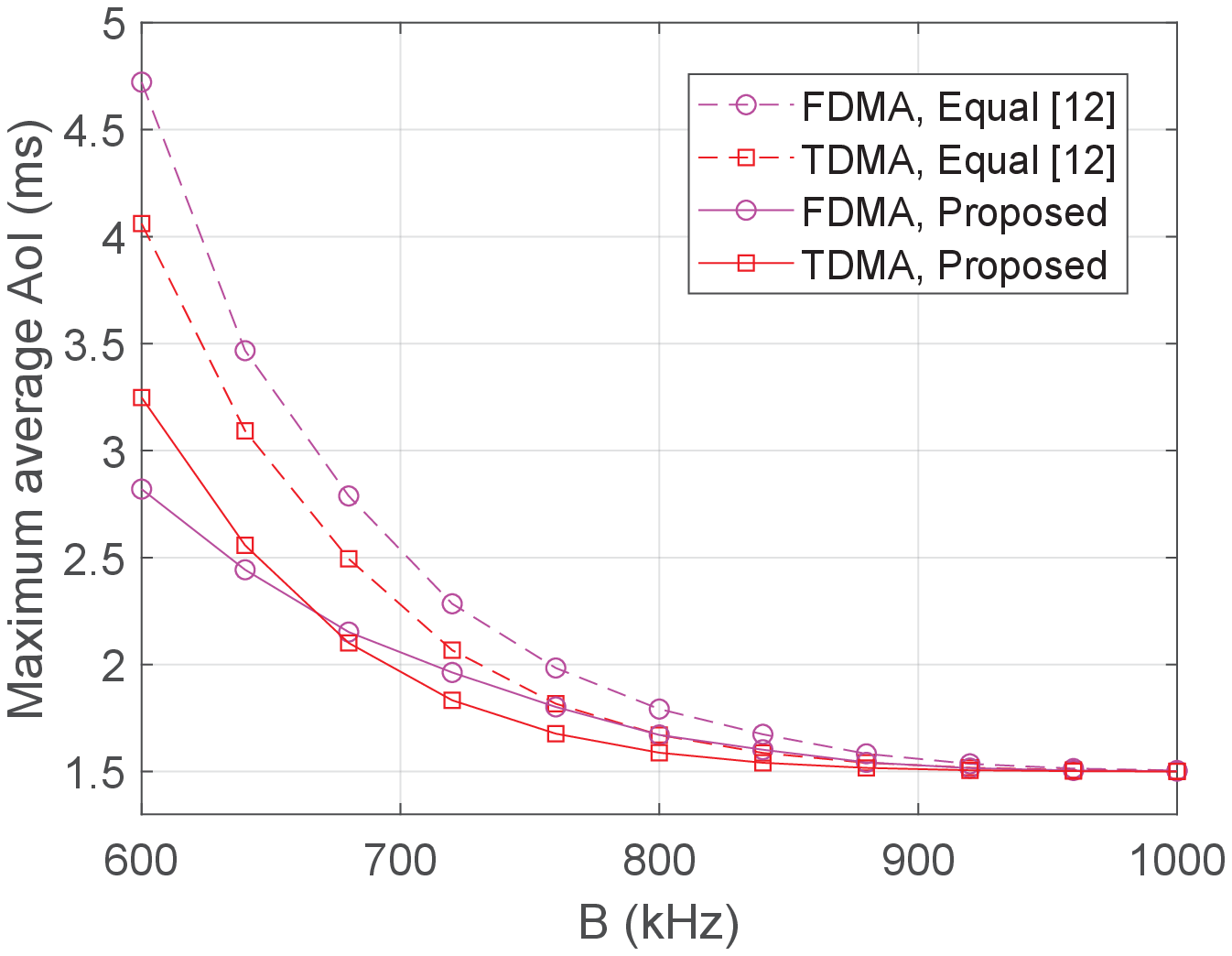}
		\end{minipage}
	} \hspace{3mm}
	\subfigure[{The maximum average AoI versus the length of information bits.}]{
		\begin{minipage}{0.3\linewidth}
			\centering
			\includegraphics[width=2.22in,height=1.93in]{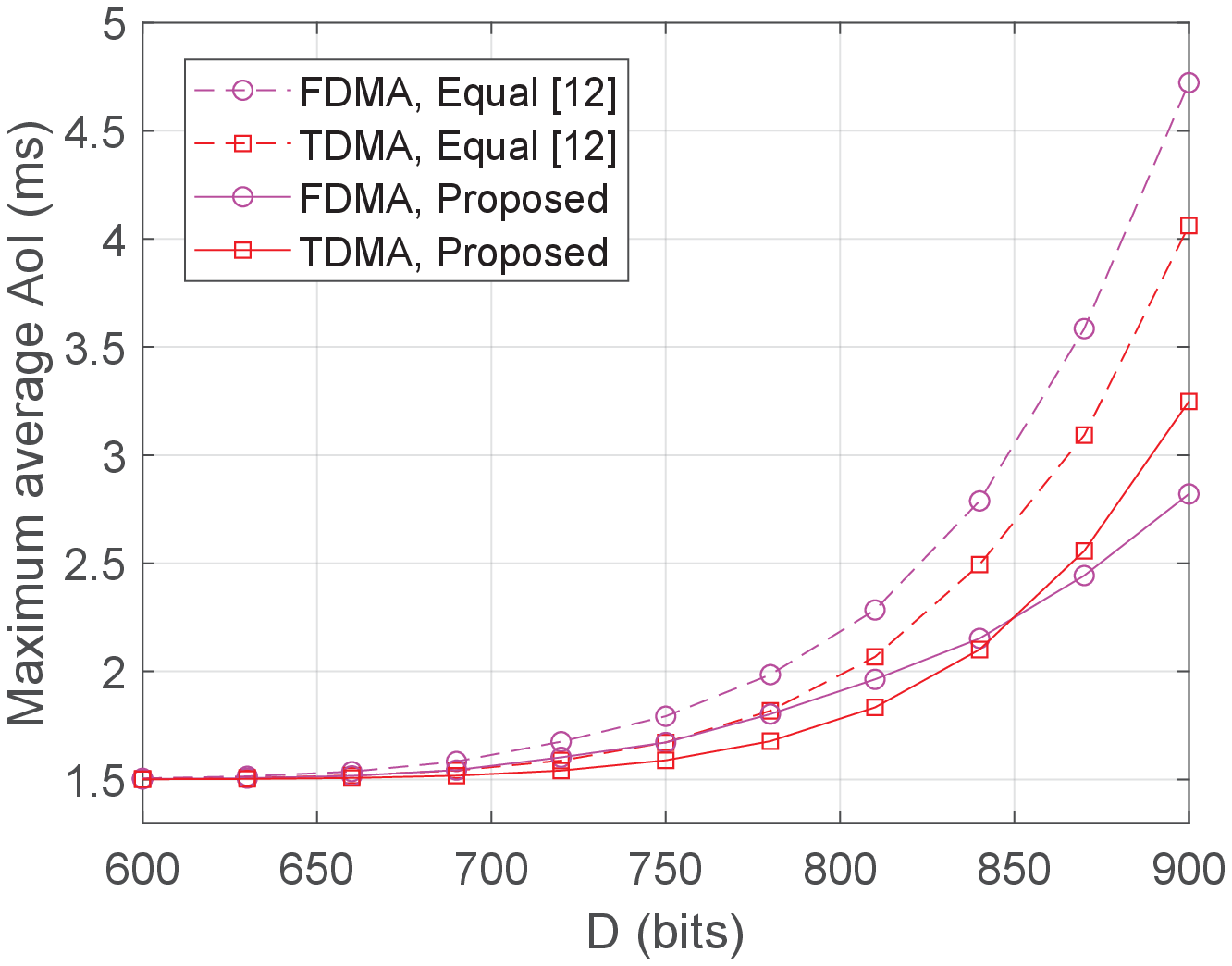}
		\end{minipage}
	}
	\caption{Performance comparison of the proposed designs and benchmarks.}
	\label{fig3}
	\vspace{0mm}
\end{figure*}

\vspace{10pt}

Fig.\ref{fig3}(a) plots the maximal average AoI versus the number of reflecting elements, $M$. It is observed that via optimizing the resource allocation, the proposed designs outperform the benchmarks in \cite{13}, wherein the time/frequence resource are equally allocated to the users. Moreover, it shows that the TDMA transmission scheme achieves lower AoI than the FDMA scheme. This is as expected because in the former case the SNR of the currently scheduled device can be maximized via dynamically adjusted IRS passive beamforming, which thus achieves lower PER for each device. It also should be noted that as the number of IRS elements increases, the performance gap between our proposed design and the benchmarks becomes smaller and then approaches zero. This is because as $M$ increases, the passive beamforming gain and thus the SNR of each packet is sufficiently large for approximately achieving error-free transmission, which renders the optimization of time/bandwidth allocation unnecessary.

Fig.\ref{fig3}(b) shows the maximum average AoI versus the total bandwidth, $B$. Similarly, it is observed the proposed designs achieve better AoI performance than the equal time/bandwidth allocation schemes as in \cite{13}. More interestingly, it shows that for the case when the total bandwidth is small, the proposed \negthinspace FDMA \negthinspace transmission scheme even outperforms the proposed \negthinspace TDMA \negthinspace one, i.e., the latter is not always more favourable as shown in Fig. 3(a). This is because when frequency resource is extremely limited, optimizing the bandwidth allocation is more efficient than dynamically adjusting the passive beamforming for improving the performance of the worst user. However, as the bandwidth increases, such performance gain becomes marginal, while the time-varying passive beamforming associated with each scheduled user is more useful, and thus the proposed \negthinspace TDMA \negthinspace transmission scheme outperforms the \negthinspace FDMA \negthinspace one. Even more, as $B$ increases up to sufficiently large (e.g., \negthinspace 1000 \negthinspace KHz as in the considered setup), both benchmark schemes achieve almost the same AoI performance as our proposed design. This is because in this case the AP has sufficient bandwidth to encode each device's packet, of which the blocklength is sufficiently long and thus the PER approaches 0, even without optimizing the time/bandwidth allocation.

Fig.\ref{fig3}(c) shows the maximum average AoI versus the length of information bits for each device, $D$. It is first observed that when the length of information bits is small, all considered four schemes achieves almost the same maximum average AoI. The reason is that the blocklength in this case is sufficient large as compared to the limited information bits, and thus almost 0 PER is achieved. However, as the length of information bits increases, owing to the optimized resource allocation, our proposed schemes outperforms the benchmark ones and the performance gap enlarges. Moreover, as $D$ increases up to larger than 850 bits, the proposed FDMA design outperforms the proposed TDMA design. This reveals that when bandwidth becomes a bottleneck for encoding such large information bits, optimizing the bandwidth allocation is more effective than dynamically adjusting the passive beamforming for decreasing the PER of the worst device.
 \vspace{0pt}

\section{Conclusions}\label{Conclusions}

In this correspondence, we investigated the downlink multi-device URLLC system assisted by IRS, wherein the FDMA and TDMA transmission schemes are compared from the perspective of maximum average AoI. Specifically, an alternating optimization based algorithm was proposed to minimize the maximal average AoI by jointly optimizing resource allocation and reflective beamforming. In the future work, a more comprehensive comparison with other typical multiple access schemes will be conducted.

\vspace{0pt}
\small{\bibliographystyle{IEEEtran}}


\end{document}